\documentclass[11pt,twoside]{article}
\PassOptionsToPackage{hyphens}{url}\usepackage{asp2014}
\aspSuppressVolSlug
\resetcounters
\usepackage{eurosym}
\newcommand{\celsius}{\mbox{$^\circ$C}}

\newcommand{\meuro}{\mbox{M\euro}}
\newcommand{\co}{\mbox{CO$_2$}}
\newcommand{\coe}{\mbox{CO$_2$e}}

\newcommand{\tcoe}{\mbox{tCO$_2$e}}
\newcommand{\ktcoe}{\mbox{ktCO$_2$e}}

\newcommand{\Mtcoe}{\mbox{MtCO$_2$e}}
\newcommand{\tcoeyr}{\mbox{tCO$_2$e yr$^{-1}$}}
\newcommand{\ktcoeyr}{\mbox{ktCO$_2$e yr$^{-1}$}}
\newcommand{\Mtcoeyr}{\mbox{MtCO$_2$e yr$^{-1}$}}
\newcommand{\moneyeft}{\mbox{tCO$_2$e \meuro$^{-1}$}}

\newcommand{\poweref}{\mbox{gCO$_2$e kWh$^{-1}$}}

\begin{document}

\title{The carbon footprint of astronomical observatories}

\author{J\"urgen~Kn\"odlseder$^1$}
\affil{$^1$, Institut de Recherche en Astrophysique en Plan\'etologie Toulouse, France; \email{jknodlseder@irap.omp.eu}}
\paperauthor{J\"urgen~Kn\"odlseder}{jknodlseder@irap.omp.eu}{ORCID}{}{GAHEC}{Toulouse}{}{31028}{France}




\begin{abstract}

The carbon footprint of astronomical research is an increasingly topical issue.
From a comparison of existing literature, we infer an annual per capita carbon footprint of several tens 
of tonnes of \co\ equivalents for an average person working in astronomy.
Astronomical observatories contribute significantly to the carbon footprint of astronomy, and we examine 
the related sources of greenhouse gas emissions as well as lever arms for their reduction.
Comparison with other scientific domains illustrates that astronomy is not the only field that needs to 
accomplish significant carbon footprint reductions of their research facilities.
We show that limiting global warming to 1.5\celsius\ or 2\celsius\ implies greenhouse gas emission 
reductions that can only be reached by a systemic change of astronomical research activities, and we
argue that a new narrative for doing astronomical research is needed if we want to keep our planet 
habitable.

\end{abstract}

\section{Introduction}

Like many years in the recent past, 2022 was another example for record-breaking climate extremes 
that have made the headlines.
Heat waves with temperatures exceeding 40\celsius\ for prolonged periods hit many places, leading 
to wildfires that were especially devastating this year in Spain and Portugal \citep{pratt2022}.
At the same time, heavy rainfall led to widespread flooding that cost thousands of lives and created
millions of displacements, this year particularly disastrous in Pakistan and Nigeria 
\citep{trenberth2022}.
While some countries flood, other regions are parched.
The Greater Horn of Africa is bracing for a fifth consecutive failed rainy season, with over 50 million 
people suffering from acute food security \citep{icpac2022}.

These events are examples of observed impacts from climate change, as summarised in the sixth assessment 
report of working group II of the Intergovernmental Panel on Climate Change (IPCC) \citep{ipcc2022wg2}.
According to the report, increased global warming will ``cause unavoidable increases in multiple climate 
hazards and present multiple risks to ecosystems and humans''.
Recognising these risks, 192 countries have signed the Paris Climate Agreement that aims ``holding 
the increase in the global average temperature to well below 2\celsius\ above pre-industrial levels and 
pursuing efforts to limit the temperature increase to 1.5\celsius\ above pre-industrial levels''.
In order to reach this goal, the agreement prescribes ``to achieve a balance between anthropogenic 
emissions by sources and removals by sinks of greenhouse gases in the second half of this century''.
In other terms, humanity needs to reach net-zero emissions by 2050, i.e.~in less than 30 years
from now.

Near-term greenhouse gas (GHG) emission reduction goals are country specific.
For example, the European climate law creates a legal obligation to reduce by 2030 GHG emissions
in the European Union by at least 55\%, compared to 1990 levels \citep{ec2021}.
The United States have set a comparable target of 50-52\% below 2005 levels in 2030 \citep{us2022}. 
Other high-income countries have declared similar targets.
Adjusting from the reference years' to today's GHG emissions, these targets translate into reductions 
of about $\sim40\%$ between today and 2030, corresponding to reductions of about $\sim7\%$ per year.
This level is comparable to, but slightly larger than, the emission reductions that were reported due to 
COVID-19 lockdowns in 2020 \citep{jackson2022}.
And this level needs to be achieved continuously in high-income countries over the next three decades, 
with each year's reduction coming on top of those achieved in the year before.

In their sixth assessment report, the working group III of the IPCC emphasises that limiting warming to 
1.5\celsius\ or 2\celsius\ ``involve[s] rapid and deep and in most cases immediate GHG emission 
reductions in all sectors'' \citep{ipcc2022wg3}.
Astronomy, as we will see, is a sector that contributes to global warming, and consequently it is concerned
by this imperative.
The IPCC reminds that ``doing less in one sector needs to be compensated by further reductions in other 
sectors if warming is to be limited'' \citep{ipcc2022wg3}.
Eventually, policy makers and the society at large may set different reduction efforts for different activity
sectors, yet until this is done, a reasonable assumption is that astronomy needs to reduce GHG emissions 
at a typical rate of $\sim7\%$ per year over the next three decades.

\section{The carbon footprint of astronomy}

In view of the challenge described above, several astronomical institutes and communities have 
assessed their carbon footprints with the aim to identify lever arms for GHG emission reductions.
\citet{stevens2020} estimated the GHG emissions of Australian astronomers, including 
powering office buildings,
business flights, 
supercomputer usage, 
and electricity consumption for operations of some ground-based observatories.
\citet{jahnke2020} estimated the GHG emissions of astronomers at the Max Planck Institute for 
Astronomy (MPIA) in Germany, including
business flights, commuting, electricity, heating, computer purchases, paper use, and cafeteria meat 
consumption.
\citet{vandertak2021} estimated the GHG emissions arising from professional astronomy activities in 
the Netherlands, including
business flights, commuting, electricity, heating and supercomputing.
\citet{martin2022} estimated the GHG emissions of the Institut de Recherche en Astrophysique et 
Plan\'etologie (IRAP), which is the largest astrophysics laboratory in France, including
running the office buildings (electricity, heating, water, air conditioning, waste),
food,
commuting,
professional travelling,
purchase of goods and services,
external computing and
use of observational data.
The later work is the most comprehensive carbon footprint estimate performed to date, and the only study 
that employed a formal carbon accounting methodology.

\articlefigure[width=\textwidth]{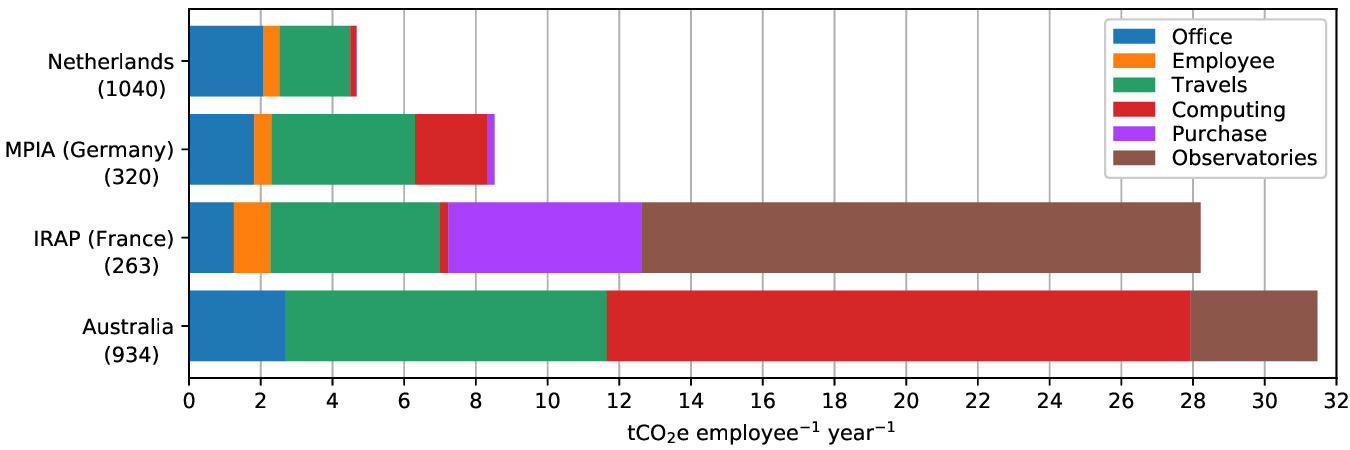}{fig:astrofootprint}{Literature estimates of the annual average
carbon footprint of a person working in astronomy. The numbers indicate the staff personal used to 
compute the per capita footprint.}

The results of these studies, expressed as annual per capita GHG emissions for a person working
at an astronomy institute, are summarised in Fig.~\ref{fig:astrofootprint}.
The results were grouped by emission sources, where ``Office'' combines all GHG emissions
related to running the office building (electricity, heating, water, air conditioning, waste) and ``Employee'' 
covers food and commuting.
\citet{jahnke2020} presented a similar figure to compare the MPIA carbon footprint to the one of Australia,
and like for their comparison, we multiplied the air travel footprint quoted by \citet{stevens2020} for 
Australia by a factor of two to account for non-\co\ radiative forcing of the same level as in all other
studies.
In their Fig.~1, \citet{jahnke2020} compared the carbon footprint per researcher despite the fact that the 
definition of ``researcher'' differed among the studies (PhD students were included for Australia but
excluded for MPIA).
A more relevant denominator is the total number of employees since they all contribute to the carbon 
footprint of a given research institute, and consequently, we use the carbon footprint per employee 
for our comparison.

The most striking feature of Fig.~\ref{fig:astrofootprint} are the huge differences between the various
carbon footprint estimates, which are partially explained by the carbon intensity of electricity
production in a given country.
For example, the largest contributor to the GHG emissions in Australia is computing with 
16.2 \tcoeyr\ per employee.
Would the same amount of computing be performed in France, which has a carbon intensity
for electricity production that is about 14 times lower than in Australia, computing would only 
account for 1.2 \tcoeyr\ per employee.
The relatively modest office footprint in Australia with respect to the other countries comes therefore as
surprise, since with the aforementioned large differences in carbon intensity we would expect an
office footprint that is one order of magnitude larger for Australia compared to France.
Note that \citet{jahnke2020} used in their study for MPIA the quoted carbon intensity on the electricity bill, 
and \citet{vandertak2021} took into account green certificats for electricity purchase, putting the
related GHG emissions to zero.
Yet even with green certificats, electricity production is not emission free, and it has been argued
that discounting on the basis of green certificats fails to provide accurate or relevant information in 
GHG reports \citep{brander2018}.
Consequently, carbon accounting methods either prescribe discounting of green certificats 
(Bilan Carbone\textsuperscript{\textcopyright})
or impose dual reporting (GHG protocol, Carbon Disclosure project).
Using the so called location-based method that applies grid average carbon intensities to electricity
consumption, and that were used by \citet{martin2022} for IRAP and \citet{stevens2020} for Australia,
would have led to larger office-related footprints for MPIA and the Netherlands.

The travel footprint, which is largely dominated by air travelling, varies among the different studies
with the smallest footprint of $\sim2$ \tcoeyr\ per employee estimated for the Netherlands and
the largest footprint of $\sim9$ \tcoeyr\ per employee estimated for Australia.
While the latter can plausibly be explained by the size and remote location of Australia
\citep{stevens2020}, the former may be related to the large number of PhD students that are 
among the employees in the Netherlands, as studies have revealed that students travel less
compared to senior scientists \citep[e.g.][]{pargman2022}.

Most of the studies ignored some potentially important GHG emission sources (or considered 
them only partially), such as purchase of goods and services and use of observational data.
Only the IRAP estimate provides a comprehensive assessment of all astronomy-related GHG
emissions \citep{martin2022}, resulting in a per-capita footprint of $\sim28$ \tcoeyr.
Since France has with $60$ \poweref\ a rather low carbon intensity for electricity production
\citep[compared to $415$ \poweref\ for the Netherlands, $460$ \poweref\ for Germany, 
and $840$ \poweref\ for Australia, see][]{basecarbone}, the IRAP estimate presents likely a lower 
limit to the average carbon footprint of a person working in astronomy.
Adding the IRAP estimates for purchase and use of observational data on top of the other estimates 
in Australia would lead to a per-capita footprint of about $\sim50$ \tcoeyr, which may be typical
for a country that has a large carbon intensity for electricity production and that relies heavily
on air travelling.
While reducing the carbon intensity for electricity production (for example by switching to solar
panels or wind turbines) and reducing air travelling (for example by using videoconferencing)
will have a direct impact on the carbon footprint of astronomy, the lever arms to reduce the 
footprint of astronomical observatories are less obvious.
We therefore will examine the carbon footprint of astronomical observatories now in more detail.

\section{The carbon footprint of astronomical observatories}

The annual carbon footprint of astronomical observatories was assessed by \citet{knodlseder2022}
to $1.2\pm0.2$ \Mtcoeyr\ for the year 2019, which is the reference year for all values quoted in
this section.
To derive this estimate, the authors used a monetary method that relates construction and
operating costs for ground-based observatories and mission cost for space-based facilities
to GHG emissions, complemented by an assessment based on payload launch mass for
space missions.
Applying this method to a list of 85 astronomical facilities, and extrapolating the results to the
world-fleet of astronomical observatories led to the annual carbon footprint quoted above.
By comparing the number of authors from IRAP to all authors in the world that published
refereed articles citing a given astronomical facility, \citet{martin2022} attributed a footprint 
of 4.1 \ktcoeyr\ to IRAP, corresponding to $0.3\%$ of the total annual carbon footprint of
all facilities.
This results in a per-capita footprint of $15.6$ \tcoeyr\ related to astronomical observatories
at IRAP.
Dividing the total annual carbon footprint of all facilities by this per-capita footprint suggests
that around $77,000$ people are working in astronomy world-wide, which is about twice
the number of $30,000$ astronomers with PhD degree in the world, as estimated by 
\citet{knodlseder2022}.

To estimate the carbon footprint of space missions, \citet{knodlseder2022} derived a 
carbon intensity of 140 \tcoe\ per \meuro\ of mission cost from the literature.
Compared to carbon intensities for other activity sectors this is a relatively small value, reflecting
the fact that a non-negligible fraction of the GHG emissions from the space sector are arising from
office work, which has a relatively small carbon intensity compared to other, more material-intensive
activities \citep{chanoine2017}.
Consequently, decarbonising the powering and heating of office buildings is an important lever
arm for carbon footprint reductions of space facilities.

It is estimated that between 50\% to 70\% of the carbon footprint of a space mission is related 
to the launcher \citep{maury2019}, putting the decarbonisation of the launcher in the focus 
of the space sector.
\citet{harris2019} estimate the cradle-to-use carbon footprint of a Falcon 9 rocket and a Falcon 
Heavy rocket to about $\sim10$ \ktcoe\ and $\sim15$ \ktcoe, respectively.
Owing to the large carbon footprint of propellant production for an Ariane 5 rocket, and based on
information published by \citet{chanoine2017} and \citet{schabedoth2020}, we estimate its 
footprint to about twice that of a Falcon Heavy rocket.
Dividing these footprints by the launcher costs results in a comparable carbon intensity of about
$\sim180$ \moneyeft\ for all three launchers, suggesting that a reduction of launch cost
leads to a reduction of carbon footprint.
Given that the reduction of launch costs is pursued to facilitate access to space, it is likely that
the per launcher footprint reductions will be quickly outpaced by the increasing number of 
launches, suggesting the appearance of a classical rebound effect that will prevent a reduction in GHG 
emissions in the space sector.
Probably more promising is the decarbonisation of propellant production, as planned for example
by the Hyguane project for Europe's spaceport Kourou in French Guiana \citep{esa2022}.
Using methane as alternative propellant for launchers also provides potential for carbon
footprint reductions \citep{kurela2020}.

On top of the launcher footprint comes the carbon footprint of the space centres from which the 
rockets are launched, for example $\sim100$ \ktcoeyr\ for the Kennedy Space Center 
\citep{nasa2020} and $\sim40$ \ktcoeyr\ for Kourou \citep{cnes2019}, with electricity
consumption being the most significant source of GHG emissions in both cases.
Additional contributions come from the payload production, including the production of the
astronomical relevant probes or telescopes, the test campaigns and the associated infrastructures,
integration activities, and ground operations following the launch.
For example, \citet{sydnor2011} estimate the carbon footprint of NASA's ground test facilities
in 2008-2010 to 137 \ktcoeyr, with electricity consumption and natural gas being the most
significant contributors.
For all these contributions, moving away from fossil fuels towards renewable energies will have 
the potential to considerably reduce the GHG emissions.

Combining all contributions, aggregated lifecycle carbon footprints for space missions range 
from a few tens of \ktcoe\ for Small Explorer (SMEX) missions like the Nuclear Spectroscopic 
Telescope Array (NuStar) or the Galaxy Evolution Explorer (GALEX) to over one million tonnes 
of \coe\ for flagship missions like the James Webb Space Telescope (JWST) \citep{knodlseder2022}.
While decarbonising electricity production for all lifecycle stages of a space mission has a
significant decarbonisation potential, the current increase in the number of new space missions,
favoured by the ``newspace'' boom that accompanies the transition from government-funded
to commercial space activities, has the potential to annihilate the efforts.

The situation is not much different for ground-based observatories.
\citet{knodlseder2022} estimated carbon intensities for construction and operations of
ground-based observatories to 240 \moneyeft\ and 250 \moneyeft, respectively.
While detailed lifecycle analyses for ground-based observatories are still lacking, according to 
\citet{eso2019}, who assessed the carbon footprint of the European Extremely Large Telescope
(E-ELT), the construction GHG emissions seem to arise primarily from concrete and steel (like in 
a typical building construction project), as well as from installations, mechanisms and freight.
For operations, GHG emissions arise primarily from electricity consumption, followed by purchase
of goods and services, business travels, commuting and freight \citep{eso2019}.
Options for carbon footprint reductions include use of low carbon concrete, recycled steel and
bio sourced materials such as wood for construction, while a shift to renewable energies can
help to decarbonise operations.
Yet these decarbonisation efforts will only be effective if no additional sources of GHG emissions
are added.

\articlefigure[width=\textwidth]{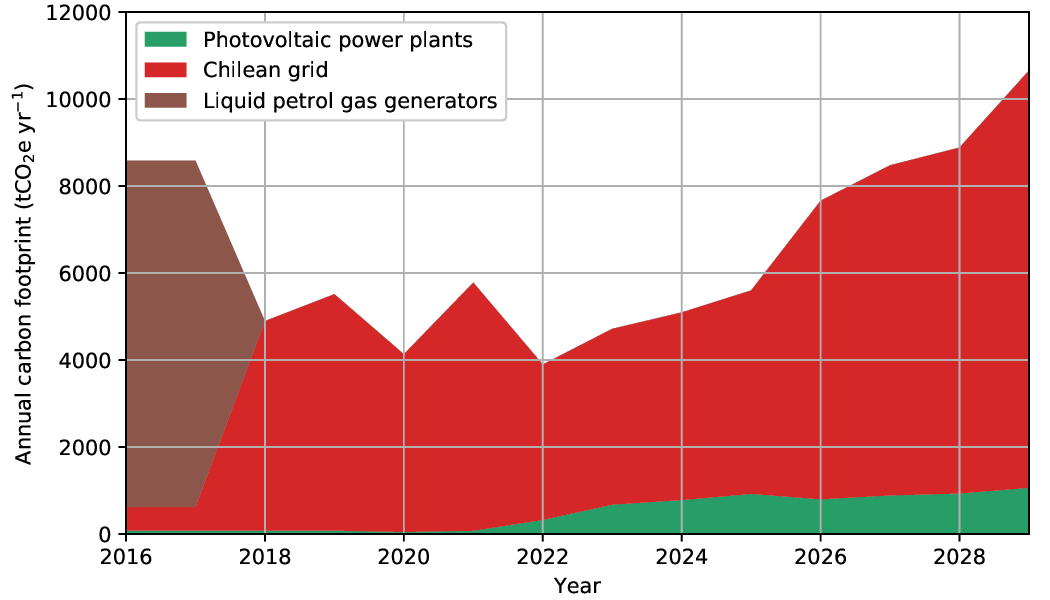}{fig:esofootprint}{Past and predicted annual carbon 
footprint of electricity consumption at the ESO observatory sites in La Silla, Paranal and 
Armazones \citep[data from][]{filippi2022}.}

This is illustrated in Fig.~\ref{fig:esofootprint} that shows the past and predicted annual 
carbon footprint of electricity consumption at the ESO observatory sites in Chile.
While between 2016 and 2022 a reduction of GHG emissions from electricity consumption by 
$\sim50\%$ was achieved (by swapping at Paranal from liquid petrol gas generators to a grid connection 
in 2018 and adding photovoltaic power plants in 2022), the additional electricity needs of E-ELT will 
have annihilated all the reductions by the end of this decade; despite important efforts, the GHG 
emissions due to electricity consumption will exceed in 2030 those of 2016.
This illustrates an obvious but inconvenient truth: it is extremely difficult to decarbonise while ramping
up!
ESO is so far the only organisation that provides public information on carbon footprint estimates and 
reduction plans, exposing the organisation obviously to be used as a case-study.
There are no reasons to believe that the situation is different for other organisations, at least as long as 
they continue to expand.

\section{Other scientific domains}

\articlefigure[width=\textwidth]{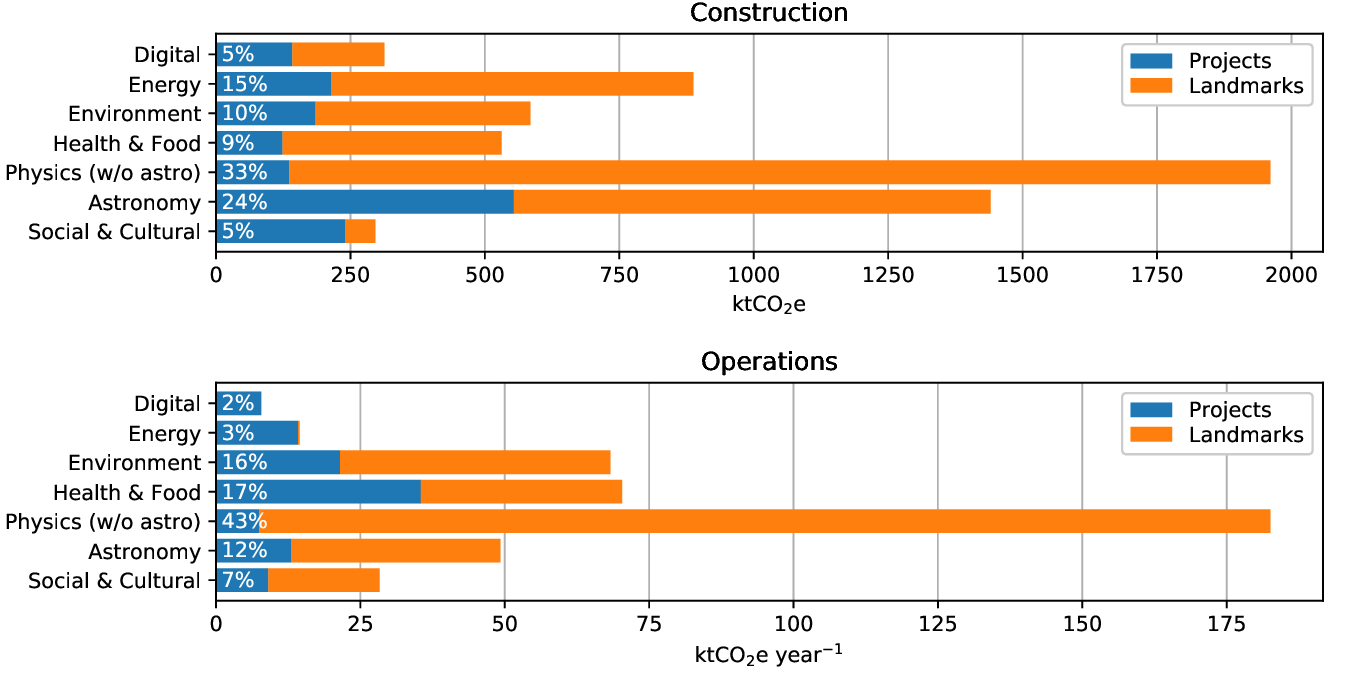}{fig:esfrifootprint}{Estimated construction (top) and
annual operations (bottom) carbon footprint of projects and landmarks in the ESFRI 2021
infrastructure roadmap, aggregated by scientific domain \citep[data from][]{esfri2021}.
Percentages give the contribution of a scientific domain to the total carbon footprint.}

The situation for astronomy is now clear, but what's the situation for other scientific domains that rely
heavily on large-scale research infrastructures?
To shed some light on this question, estimates of the carbon footprint of research infrastructures
that are currently under development in Europe in different scientific domains are shown in 
Fig.~\ref{fig:esfrifootprint}.
We based the figure on the 2021 update of the ESFRI infrastructure roadmap \citep{esfri2021},
and distinguished infrastructures in the preparation phase (called ``projects'' in the ESFRI
roadmap) and infrastructures that were implemented or reached an advanced implementation 
phase (called ``landmarks'').
We estimated the carbon footprints by converting construction and operating costs as quoted
in the ESFRI roadmap into GHG emissions using the same emission factors that were derived 
by \citet{knodlseder2022} for ground-based astronomical observatories.
For the latter, we assume that all infrastructures listed in the ESFRI roadmap will be operating
at some point contemporaneously, so that their footprints will add up.
For comparison of astronomy with other scientific domains, we split the physics domain that 
includes astronomy in the ESFRI roadmap into astronomy infrastructures and other physics
infrastructures.

According to this analysis, the total carbon footprint for construction of planned research 
infrastructures in Europe amounts to about $\sim6$ \Mtcoe\ with an expected associated annual 
operations footprint of $\sim420$ \ktcoeyr, suggesting that the annual operations footprint amounts
to $\sim7\%$ of the construction footprint \citep[for comparison, the annual GHG emissions
of Europe in 2019 amounted to about $\sim4000$ \Mtcoe, see][]{eaa2022}.
The construction of astronomical research infrastructures contribute to $24\%$ to the total construction 
carbon footprint in Europe, while for operations the contribution is estimated to $12\%$.
Research infrastructures in the domain of physics account for $33\%$ of the construction footprint,
raising to $57\%$ when astronomy is included.
The share in the combined physics operations footprint is comparable, suggesting that research
infrastructures for physics account for more than half of the total carbon footprint in Europe.
Examples of physics infrastructures with a large estimated construction carbon footprints include the 
European Spallation Source ($\sim720$ \ktcoe), the European X-Ray Free-Electron Laser Facility
($\sim370$ \ktcoe) and the High-Luminosity Large Hadron Collider ($\sim340$ \ktcoe).
All three facilities have estimated annual operation footprints of $\sim35$ \ktcoeyr, in excess
of those of the largest astronomical facilities \citep{knodlseder2022}.
Obviously, astronomy is not the only domain that has an important GHG emission reduction
burden, particle physics is certainly in a similar situation.
An assessment of the global carbon footprint of particle physics is still lacking, yet some first estimates
of specific GHG emission sources are appearing in the literature \citep{bloom2022,janot2022}.

\section{Towards sustainable astronomy}

Coming back to astronomy, the question we have to face now in our field is how we can bring our 
activities in line with the imperative to reduce GHG emissions at a rate of $7\%$ per year over the 
next three decades.
Or to be more specific: what will we change in 2023 to reduce GHG emissions by $7\%$ next year?
The required amplitude of GHG emission reductions is significant, and they are not achievable by
fiddling on the margins.
Reductions of $7\%$ are superior to what was achieved following the COVID-19 lockdowns in 2020
which had important impacts on human activities with still visible economic aftermaths today.
And as these aftermaths illustrate, reductions of $7\%$ are not possible without systemic changes,
giving up the way we operate currently and inventing a new narrative for doing astronomical research 
in the future.

A first step would be to pull the breaks and focus on using what we already have instead of
aiming for new facilities.
Implementing technical decarbonisation solutions will take some time and will need investments,
making them solutions for the mid- or long-term, yet unlikely for the immediate future.
An example of short-term solutions would be to shift activities away from building new research
infrastructures towards a more extended and deeper exploitation of existing facilities, and
the analysis of existing astronomical data archives.
It is well recognised that archives are valuable resources for astronomy, and a significant
fraction of discoveries is made by exploring already existing data \citep[e.g.][]{white2009,demarchi2022}.
Use of archival data shall be actively promoted and be considered the baseline for
astronomical research.

A next step would be to base resource allocation decisions on their associated carbon footprint, having 
in mind that remaining carbon budgets that keep global warming below 1.5\celsius\ or 2\celsius\ 
shrink rapidly.
Today, no funding agency is investing significantly into decarbonising research infrastructures;
tomorrow, decarbonising existing facilities must become their funding priority!
This also means that less money will be available to build new infrastructures, yet is this really
a problem?
\citet{stoehr2015} argue that, in the future, observatories will compete for astronomers to work with 
their data, which if true seems to indicate that we may have already too many facilities.
There is no requirement {\em per se} on the deployment pace of new facilities or missions, and
slowing down the current pace will lead to less GHG emissions, free resources for investing into 
decarbonisation and give more time for in-depth analyses of existing data.

Another measure is moving away from competition towards more collaboration.
If we really believe that astronomers are working for mankind, there is no need to build the same kind
of facility several times on the globe.
For example, one 40-m class telescope in the world should be sufficient to make the discoveries to 
be made with such an instrument.
And there is no scientific justification for having a new space race towards the Moon or the planets, 
a few well-coordinated international missions should be sufficient to gain the knowledge we are after.

Of course, astronomy is not the root cause of climate change, nor can astronomy alone fix it,
but astronomy with its significant per capita GHG emissions must be exemplary and take its fair share,
leading the way towards a sustainable future for humanity on Earth.
It's time to act now, before we are forced in an uncontrolled and likely very violent way to respect the 
boundaries of our planet.

\begin{acknowledgements}
The author would like to thank 
R. Arsenault,
S. Brau-Nogu{\'e},
M. Coriat,
P. Garnier,
A. Hughes,
P. Martin,
A. Stevens and
L. Tibaldo for useful discussions.
This work has benefited from discussions within the GDR Labos~1point5.
\end{acknowledgements}

\bibliography{I04}

\begin{thebibliography}{}
\expandafter\ifx\csname natexlab\endcsname\relax\def\natexlab#1{#1}\fi
\expandafter\ifx\csname url\endcsname\relax
  \def\url#1{\texttt{#1}}\fi
\expandafter\ifx\csname urlprefix\endcsname\relax\def\urlprefix{URL }\fi
\providecommand{\eprint}[2][]{\url{#2}}

\bibitem[{Bloom et~al.(2022)Bloom, Boisvert, Britzger, Buuck, Eichhorn,
  Headley, Lohwasser, \& Merkel}]{bloom2022}
Bloom, K., Boisvert, V., Britzger, D., Buuck, M., Eichhorn, A., Headley, M.,
  Lohwasser, K., \& Merkel, P. 2022, {Climate impacts of particle physics}.
  \urlprefix\url{https://arxiv.org/abs/2203.12389}

\bibitem[{Brander et~al.(2018)Brander, Gillenwater, \& Ascui}]{brander2018}
Brander, M., Gillenwater, M., \& Ascui, F. 2018, Energy Policy, 112, 29.
  \urlprefix\url{https://www.sciencedirect.com/science/article/pii/S0301421517306213}

\bibitem[{Breitenstein(2021)}]{basecarbone}
Breitenstein, A. 2021, {Base Carbone}, \url{https://bilans-ges.ademe.fr/en}

\bibitem[{{Chanoine} et~al.(2017){Chanoine}, {Duvernois}, \& {Le
  Guern}}]{chanoine2017}
{Chanoine}, A., {Duvernois}, P.-A., \& {Le Guern}, Y. 2017, {Environmental
  impact assessment analysis}, ESA

\bibitem[{{De Marchi}(2022)}]{demarchi2022}
{De Marchi}, G. 2022, in American Astronomical Society Meeting Abstracts,
  vol.~54 of American Astronomical Society Meeting Abstracts, 302.13

\bibitem[{{EC}(2021)}]{ec2021}
{EC} 2021, Regulation (eu) 2021/1119 of the european parliament and of the
  council of 30 june 2021 establishing the framework for achieving climate
  neutrality and amending regulations (ec) no 401/2009 and (eu) 2018/1999
  (`european climate law'),
  \url{https://eur-lex.europa.eu/legal-content/EN/TXT/?uri=CELEX:32021R1119}.
  Accessed: 2022-10-21

\bibitem[{{EEA}(2022)}]{eaa2022}
{EEA} 2022, Annual european union greenhouse gas inventory 1990-2020 and
  inventory report 2022,
  \url{https://www.eea.europa.eu//publications/annual-european-union-greenhouse-gas-1}

\bibitem[{{ESA}(2022)}]{esa2022}
{ESA} 2022, Launch goes green with spaceport hydrogen plan,
  \url{https://www.esa.int/Enabling_Support/Space_Transportation/Europe_s_Spaceport/Launch_goes_green_with_Spaceport_hydrogen_plan}.
  Accessed: 2022-10-25

\bibitem[{{ESFRI}(2021)}]{esfri2021}
{ESFRI} 2021, Strategy report on research infrastructures

\bibitem[{{Filippi} et~al.(2022){Filippi}, {Scibior}, {van der Heyden},
  {Arsenault}, \& {Tamai}}]{filippi2022}
{Filippi}, G., {Scibior}, P., {van der Heyden}, P., {Arsenault}, R., \&
  {Tamai}, R. 2022, in Ground-based and Airborne Telescopes IX, edited by H.~K.
  {Marshall}, J.~{Spyromilio}, \& T.~{Usuda}, vol. 12182 of Society of
  Photo-Optical Instrumentation Engineers (SPIE) Conference Series, 121823Z

\bibitem[{{Gallo \& Bachelet}(2019)}]{eso2019}
{Gallo \& Bachelet} 2019, {Carbon footprint assessment ESO}

\bibitem[{Harris \& Landis(2019)}]{harris2019}
Harris, T.~M., \& Landis, A.~E. 2019, in 2019 IEEE Aerospace Conference, 1

\bibitem[{{ICPAC}(2022)}]{icpac2022}
{ICPAC} 2022, The greater horn of africa is bracing for a 5th consecutive
  failed rainy season,
  \url{https://www.icpac.net/news/the-greater-horn-of-africa-is-bracing-for-a-5th-consecutive-failed-rainy-season}.
  Accessed: 2022-10-21

\bibitem[{IPCC(2022{\natexlab{a}})}]{ipcc2022wg2}
IPCC 2022{\natexlab{a}}, in Climate Change 2022: Impacts, Adaptation, and
  Vulnerability. Contribution of Working Group II to the Sixth Assessment
  Report of the Intergovernmental Panel on Climate Change, edited by H.-O.
  P\"ortner, D.~Roberts, M.~Tignor, E.~Poloczanska, K.~Mintenbeck,
  A.~Alegr\'ia, M.~Craig, S.~Langsdorf, S.~L\"oschke, V.~M\"oller, A.~Okem, \&
  B.~Rama (Cambridge, UK and New York, NY, USA: Cambridge University Press)

\bibitem[{IPCC(2022{\natexlab{b}})}]{ipcc2022wg3}
--- 2022{\natexlab{b}}, in Climate Change 2022: Mitigation of Climate Change.
  Contribution of Working Group III to the Sixth Assessment Report of the
  Intergovernmental Panel on Climate Change, edited by P.~Shukla, J.~Skea,
  R.~Slade, A.~A. Khourdajie, R.~van Diemen, D.~McCollum, M.~Pathak, S.~Some,
  P.~Vyas, R.~Fradera, M.~Belkacemi, A.~Hasija, G.~Lisboa, S.~Luz, \& J.~Malley
  (Cambridge, UK and New York, NY, USA: Cambridge University Press)

\bibitem[{Jackson et~al.(2022)Jackson, Friedlingstein, Qu\'er\'e, Abernethy,
  Andrew, Canadell, Ciais, Davis, Deng, Liu, Korsbakken, \&
  Peters}]{jackson2022}
Jackson, R.~B., Friedlingstein, P., Qu\'er\'e, C.~L., Abernethy, S., Andrew,
  R.~M., Canadell, J.~G., Ciais, P., Davis, S.~J., Deng, Z., Liu, Z.,
  Korsbakken, J.~I., \& Peters, G.~P. 2022, Environmental Research Letters, 17,
  031001. \urlprefix\url{https://dx.doi.org/10.1088/1748-9326/ac55b6}

\bibitem[{{Jahnke} et~al.(2020){Jahnke}, {Fendt}, {Fouesneau}, {Georgiev},
  {Herbst}, {Kaasinen}, {Kossakowski}, {Rybizki}, {Schlecker}, {Seidel},
  {Henning}, {Kreidberg}, \& {Rix}}]{jahnke2020}
{Jahnke}, K., {Fendt}, C., {Fouesneau}, M., {Georgiev}, I., {Herbst}, T.,
  {Kaasinen}, M., {Kossakowski}, D., {Rybizki}, J., {Schlecker}, M., {Seidel},
  G., {Henning}, T., {Kreidberg}, L., \& {Rix}, H.-W. 2020, Nature Astronomy,
  4, 812. \eprint{2009.11307}

\bibitem[{Janot \& Blondel(2022)}]{janot2022}
Janot, P., \& Blondel, A. 2022, The European Physical Journal Plus, 137.
  \urlprefix\url{https://doi.org/10.1140%2Fepjp%2Fs13360-022-03319-w}

\bibitem[{{Kn{\"o}dlseder} et~al.(2022){Kn{\"o}dlseder}, {Brau-Nogu{\'e}},
  {Coriat}, {Garnier}, {Hughes}, {Martin}, \& {Tibaldo}}]{knodlseder2022}
{Kn{\"o}dlseder}, J., {Brau-Nogu{\'e}}, S., {Coriat}, M., {Garnier}, P.,
  {Hughes}, A., {Martin}, P., \& {Tibaldo}, L. 2022, Nature Astronomy, 6, 503.
  \eprint{2201.08748}

\bibitem[{Kurela \& Noir(2020)}]{kurela2020}
Kurela, M., \& Noir, P. 2020, in {Congr{\`e}s Lambda Mu 22 ``Les risques au
  c{\oe}ur des transitions'' (e-congr{\`e}s) - 22e Congr{\`e}s de Ma{\^i}trise
  des Risques et de S{\^u}ret{\'e} de Fonctionnement, Institut pour la
  Ma{\^i}trise des Risques} (Le Havre (e-congr{\`e}s), France).
  \urlprefix\url{https://hal.archives-ouvertes.fr/hal-03331333}

\bibitem[{Martin et~al.(2022)Martin, Brau-Nogu{\'e}, Coriat, Garnier, Hughes,
  Kn{\"o}dlseder, \& Tibaldo}]{martin2022}
Martin, P., Brau-Nogu{\'e}, S., Coriat, M., Garnier, P., Hughes, A.,
  Kn{\"o}dlseder, J., \& Tibaldo, L. 2022, {The carbon footprint of IRAP}.
  \urlprefix\url{https://arxiv.org/abs/2204.12362}

\bibitem[{Maury(2019)}]{maury2019}
Maury, T. 2019, Theses, {Universit{\'e} de Bordeaux}.
  \urlprefix\url{https://tel.archives-ouvertes.fr/tel-02275822}

\bibitem[{{Monteith}(2020)}]{nasa2020}
{Monteith}, W.~R. 2020, {Environmental Assessment for Space X Falcon Launches
  at Kennedy Space Center and Cape Canaveral Air Force Station}, FAA

\bibitem[{Pargman et~al.(2022)Pargman, Laaksolahti, Eriksson, Rob{\`e}rt, \&
  Bi{\o}rn-Hansen}]{pargman2022}
Pargman, D., Laaksolahti, J., Eriksson, E., Rob{\`e}rt, M., \& Bi{\o}rn-Hansen,
  A. 2022, Who Gets to Fly? (Singapore: Springer Singapore), 133.
  \urlprefix\url{https://doi.org/10.1007/978-981-16-4911-0_6}

\bibitem[{{Pratt}(2022)}]{pratt2022}
{Pratt} 2022, Heatwaves and fires scorch europe, africa, and asia,
  \url{https://earthobservatory.nasa.gov/images/150083/heatwaves-and-fires-scorch-europe-africa-and-asia}.
  Accessed: 2022-10-21

\bibitem[{Schabedoth(2020)}]{schabedoth2020}
Schabedoth, P.~E. 2020, Master thesis, {Norwegian University of Science and
  Technology}.
  \urlprefix\url{https://ntnuopen.ntnu.no/ntnu-xmlui/bitstream/handle/11250/2779647/no.ntnu:inspera:57317890:36141261.pdf?sequence=1}

\bibitem[{{Serfass-Denis}(2019)}]{cnes2019}
{Serfass-Denis}, A. 2019, {Rapport reglementaire BGES}, CNES

\bibitem[{{Stevens} et~al.(2020){Stevens}, {Bellstedt}, {Elahi}, \&
  {Murphy}}]{stevens2020}
{Stevens}, A. R.~H., {Bellstedt}, S., {Elahi}, P.~J., \& {Murphy}, M.~T. 2020,
  Nature Astronomy, 4, 843. \eprint{1912.05834}

\bibitem[{{Stoehr} et~al.(2015){Stoehr}, {Lacy}, {Leon}, {Muller}, \&
  {Kawamura}}]{stoehr2015}
{Stoehr}, F., {Lacy}, M., {Leon}, S., {Muller}, E., \& {Kawamura}, A. 2015, in
  Astronomical Data Analysis Software an Systems XXIV (ADASS XXIV), edited by
  A.~R. {Taylor}, \& E.~{Rosolowsky}, vol. 495 of Astronomical Society of the
  Pacific Conference Series, 69

\bibitem[{Sydnor et~al.(2011)Sydnor, Marshall, \& McGinnis}]{sydnor2011}
Sydnor, C., Marshall, T., \& McGinnis, S. 2011, in LCA XI Conference:
  Instruments for Green Futures Markets, 1

\bibitem[{{Trenberth}(2022)}]{trenberth2022}
{Trenberth} 2022, 2022's supercharged summer of climate extremes: How global
  warming and la ni\~na fueled disasters on top of disasters,
  \url{https://theconversation.com/2022s-supercharged-summer-of-climate-extremes-how-global-warming-and-la-nina-fueled-disasters-on-top-of-disasters-190546}.
  Accessed: 2022-10-21

\bibitem[{{USA}(2022)}]{us2022}
{USA} 2022, {The United States' Nationally Determined Contribution. Reducing
  Greenhouse Gases in the United States: A 2030 Emissions Target},
  \url{https://unfccc.int/sites/default/files/NDC/2022-06/United%20States%20NDC%20April%2021%202021%20Final.pdf}.
  Accessed: 2022-10-21

\bibitem[{{Van der Tak} et~al.(2021){Van der Tak}, {Burtscher}, {Zwart},
  {Tabone}, {Nelemans}, {Bloemen}, {Young}, {Wijnands}, {Janssen}, \&
  {Schoenmakers}}]{vandertak2021}
{Van der Tak}, F., {Burtscher}, L., {Zwart}, S.~P., {Tabone}, B., {Nelemans},
  G., {Bloemen}, S., {Young}, A., {Wijnands}, R., {Janssen}, A., \&
  {Schoenmakers}, A. 2021, Nature Astronomy, 5, 1195

\bibitem[{{White} et~al.(2009){White}, {Accomazzi}, {Berriman}, {Fabbiano},
  {Madore}, {Mazzarella}, {Rots}, {Smale}, {Storrie-Lombardi}, \&
  {Winkelman}}]{white2009}
{White}, R.~L., {Accomazzi}, A., {Berriman}, G.~B., {Fabbiano}, G., {Madore},
  B.~F., {Mazzarella}, J.~M., {Rots}, A., {Smale}, A.~P., {Storrie-Lombardi},
  L., \& {Winkelman}, S. 2009, in astro2010: The Astronomy and Astrophysics
  Decadal Survey, vol. 2010, P64

\end{thebibliography}


\end{document}